\documentclass[twocolumn, a4paper, prl, ]{revtex4}

\usepackage{graphicx, epsfig, fancybox}
\usepackage{amssymb,amsmath}
\usepackage{color}

 \usepackage{times}

\def \del{\partial}    
\def \hf{\tfrac{1}{2}}    

\def \ord{\mathcal{O}}

\def\lbc{\left[}    \def\rbc{\right]}

\begin{document}

\title{Interaction ramps in a trapped Bose condensate}

\author{F.~E.~Zimmer}
\author{Masudul Haque}

 \affiliation{Max Planck Institute for the Physics of Complex
 Systems, N\"othnitzer Str.~38, 01187 Dresden, Germany}

\begin{abstract}

Non-adiabatic interaction ramps are considered for trapped Bose-Einstein
condensates.  The deviation from adiabaticity is characterized through the
heating or residual energy produced during the ramp.  We find that the
dependence of the heat on the ramp time is very sensitive to the ramp
protocol.  We explain features of this dependence through a single-parameter
effective description based on the dynamics of the condensate size.

\end{abstract}

\maketitle

%
Adiabaticity is 
a ubiquitous concept in quantum dynamics.
%
With the advent of
novel non-equilibrium experimental possibilities, 
\emph{deviations} from
adiabaticity in slow parameter changes have attracted a lot of attention
\cite{finiteRateQuenches_reviews, Polovnikov_AdiabaticPertThy,
  finiteRateQuenches_residualenergy, EcksteinKollar_NJP2010,
  CanoviRossiniFazioSantoro_JSM09, Venumadhav_PRB2010, Roux_PRA2010,
  ColluraKarevski_PRL10,   BernierRouxKollath_arxiv1010}.
The question of non-adiabaticity is of fundamental interest, but also has
practical implications.
Many experimental protocols involve adiabatically changing a parameter in
order to reach a desired quantum state.  Since non-adiabatic heating can
rarely be completely avoided, it is essential to understand deviations from
adiabaticity in slow ramps.
In addition, the proposal of \emph{adiabatic quantum computation}
\cite{AdiabaticQC} raises the question of how a realistic parameter ramp in a
quantum many-particle system deviates from adiabaticity.
While the effect of quantum critical points in the ramp path has been
considered in much detail
\cite{finiteRateQuenches_reviews,Polovnikov_AdiabaticPertThy}, first
realizations of such a quantum computer will presumably be mesoscopic rather
than macroscopic, without true quantum critical points.
Understanding non-adiabatic ramps in \emph{finite} quantum systems is
therefore vital.  
A few studies of ramps in finite and trapped systems have appeared in the very
recent literature \cite{Venumadhav_PRB2010, Roux_PRA2010,
  ColluraKarevski_PRL10, BernierRouxKollath_arxiv1010}, indicating an emerging
recognition of the importance of this issue.

Thus motivated, in this work we consider non-adiabatic ramps in the most
emblematic physical system in the world of laser-cooled atoms, namely, an
interacting Bose-Einstein condensate in a harmonic trap.  We consider ramps of
the contact interaction from an initial value $U_i$ to a final value $U_f$,
occurring in time scale $\tau$.  By varying $\tau$, we interpolate between the
limits of the \emph{instantaneous quench} ($\tau=0$) and the \emph{adiabatic
  ramp} ($\tau\rightarrow\infty$).  For finite $\tau$, we study deviations
from adiabaticity through the heating $Q$, which we define as the final energy
at time $t>\tau$ minus the ground state enegy of the final Hamiltonian.  This
quantity is also called the residual energy or the excess energy
\cite{EcksteinKollar_NJP2010, CanoviRossiniFazioSantoro_JSM09,
  finiteRateQuenches_residualenergy, Venumadhav_PRB2010}, and may be thought
of as the ``friction'' due to imperfect adiabaticity \cite{Muga-JPB-2009}.

We find that the residual energy $Q$ decreases with $\tau$ as a power law
$Q\sim\tau^{-\nu}$ rather than exponentially, with the exponent $\nu$
depending on the shape of the ramp.  For certain ramp shapes, $Q(\tau)$ has
oscillations superposed on top of the power-law decay.  The oscillation
frequency is given by the breathing-mode frequency set by the harmonic trap.
In contrast to most studies of the heat function $Q(\tau)$, we have found a
simple physical description of the features of $Q(\tau)$.
A natural description of trapped dynamics is through a variational
wave function where the extent (radius) of the condensate is treated as a
time-dependent variational parameter.  
%
%
We find that such a ``radius dynamics'' description reproduces the residual
energy behavior surprisingly well.

We use the Gross-Pitaevskii (GP) description
\cite{pitaevskii-jetp13,gross-nc20} of condensates in an isotropic harmonic
trap $V_{\rm tr}(r)$.
%
The GP energy functional is
\begin{multline*}
E[\psi] ~=~ \int_{r} \, \psi^*(r) \lbc-\frac{\hbar^2}{2m}\nabla^2\rbc \psi(r) \\ ~+~
\frac{1}{2} U(t) \int_{r} \, \big|\psi(r)\big|^4 
~+~  \int_{r}\,  V_{\rm tr}(r)\, \big|\psi(r)\big|^2  \, .
\end{multline*}
Here $r$ is the radial position variable, and $\int_r\equiv{\int}d^Dr$ is the
spatial integral appropriate to the dimensionality $D$ of the system.  The
time-dependent parameter $U$ is the effective interaction strength whose
relation to the physical interaction is also $D$-dependent
(c.f.\ Ref.~\cite{Olshanii_PRL98} for 1D).
%
%
From here on we will use trap units, expressing lengths in units of trap
oscillator length and time in units of inverse trapping frequency.  The
condensate dynamics is given by the time-dependent GP equation, 
$i\frac{\partial\psi}{\partial t} =
-\hf\bigtriangledown^2\psi+\tfrac{1}{2}r^2 \psi+U(t)|\psi|^2\psi$.

The GP equation provides an excellent account of many aspects of trapped
condensate dynamics.  
%
%
%
We restrict ourselves to GP dynamics, which is already too rich to be studied
exhaustively. 
%
%
While physics beyond GP is more important in lower dimensions, we will treat
1D, 2D and 3D cases on an equal footing.  The questions we address 
are conceptually interesting irrespective of dimensionality.
We will also show that the dominant effects we encounter and analyze are quite
insensitive to $D$.

\begin{figure}
\centering
\includegraphics*[width=0.95\columnwidth]{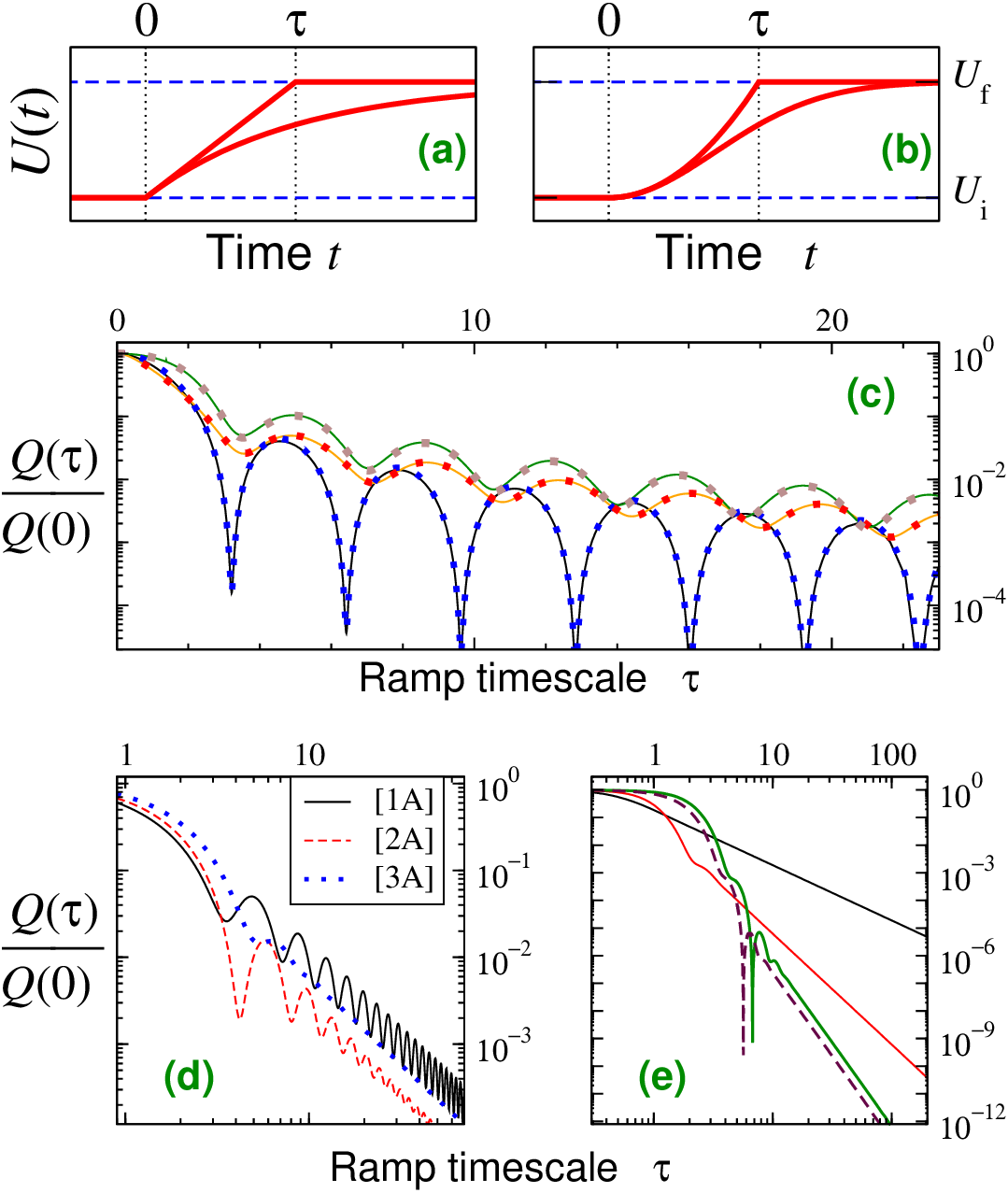}
\caption{ \label{fig_excessEnergies}
(a{\&}b)~Ramp shapes [1A], [1B], \& [2A], [2B].   
\quad (c-e)~Some residual energy curves $Q(\tau)$.
\quad (c)~For ramp shape [1A], Comparing full Gross-Pitaevskii results (full
curves) with single-parameter variational results (dotted curves).  The three
pairs have ($U_{i}$,$U_{f}$) values (0,1), (10,100), and (20,0.2),
from bottom to top.
\quad (d)~With ($U_{i}$,$U_{f}$)= (10,100), comparing the
discontinuous-derivative ramp shapes [1A], [2A], [3A].
\quad (e)~With ($U_{i}$,$U_{f}$)= (20,0.2), comparing $Q(\tau)$ for
smoothed ramp shapes, [1B], [2B], [3B], from smaller to larger slopes.
}
\end{figure}


\emph{Ramp shapes} ---
We analyze 
ramps of the form
\[
U(t) ~=~  U_{i} ~+~  \theta(t)\, (U_{f}-U_{i})\,
r(t/\tau)  \, .
\]
The ramp function $r(x)$ starts at $r(0)=0$ and ends at $r(\infty)=1$.  We do
not require $r(x>1)=1$, i.e., the ramps take place over time scale $\tau$ but
do not necessarily end at $t=\tau$.  (Contrast,
e.g., Ref.~\cite{EcksteinKollar_NJP2010}.)

Specifically, we consider the following forms for $r(x)$:
%
%
\begin{align*}
&{\rm [1A]} \quad  x\; \theta(1-x)  +\theta(x-1)  & {\rm [1B]} \quad  & 1-e^{-x} \\
&{\rm [2A]} \quad  x^2\; \theta(1-x)  +\theta(x-1)  & {\rm [2B]} \quad & 1-e^{-x^2} \\
&{\rm [3A]} \quad  x^3\; \theta(1-x) + \theta(x-1) & {\rm [3B]} \quad  & 1-e^{-x^3} 
\end{align*}
Each [A], [B] pair has the same initial behavior, $r(x){\sim}x^{\alpha}$, but
the [B] versions have no endpoint kinks.  Ramp shapes [1A] and [1B] ([2A] and
[2B]) are compared in Figure \ref{fig_excessEnergies}a
(\ref{fig_excessEnergies}b).


\emph{Residual energy features} ---
In Figure \ref{fig_excessEnergies}c-e we present the behavior of the heat
function $Q(\tau)$, normalized against its instantaneouus-quench value
$Q(\tau=0)$.  

Figures \ref{fig_excessEnergies}c shows $Q(\tau)/Q(0)$ for several ramps of
type [1A] (linear quench).  The heating displays oscillations with $\tau$, on
top of a power-law decay.  The frequency of these oscillations is the same as
the frequency of breathing-mode oscillations in real time.  Also shown, in
each case, is $Q(\tau)/Q(0)$ found from a single-parameter variational ansatz
where the cloud radius is the only variable.  The near-perfect agreement
indicates that the physics of heating in interaction ramps is almost
completely described by the radius dynamics.  In the rest of the article, we
therefore present results and analysis mostly based on the variational
description.

In Figure \ref{fig_excessEnergies}d, the residual energy curves are compared
for the ramp shapes $r(x){\sim}x^{\alpha}$ with discontinuos derivatives at
endpoints, [1A], [2A], [3A].  Each curve has an overall power-law decay
\emph{with the same decay exponent}, $Q(\tau)\sim\tau^{-2}$.  This suggests
that the residual energy for such ramps is primarily set by the endpoint kink.
%
%
Superposed on the power-law decay are oscillations, more prominent for the
linear quench $\alpha=1$ and barely visible for $\alpha=3$.

In Figure \ref{fig_excessEnergies}e we focus on smoothed ramps [1B], [2B],
[3B], which lead to non-oscillating decay of the residual energy.  The decay
exponent is seen to depend on the power $\alpha$ of $r(x)\sim{x^\alpha}$,
namely, $Q(\tau)\sim\tau^{-2\alpha}$.  

The dimensionality does not affect the decay exponents.
The $Q(\tau)$ data shown in Figures \ref{fig_excessEnergies}c-e are for $D=1$,
except for the dashed curve in \ref{fig_excessEnergies}e, which is for $D=3$
and ramp [3B].  Comparison with the corresponding $D=1$ solid curve
demonstrates that the $Q(\tau)$ behavior is practically identical in different
dimensions.

\emph{Single-parameter variational description} ---
We formulate the radius description in terms of a Gaussian variational ansatz,
which for 1D is  
\begin{equation}
\psi(x,t)=\frac{1}{[\sqrt{\pi}\sigma(t)]^{1/2}}\exp
\left[
-\frac{x^2}{2[\sigma(t)]^2}-i\beta(t) x^2
\right].
\label{eq:Ansatz1}
\end{equation}
For $D>1$ the variational wave function is a product of one such Gaussian
factor for each dimension.  Using this ansatz in the GP Lagrangian, 
we get the evolution equations for the variational parameters  $\sigma(t)$ and $\beta(t)$ 
\cite{PerezGarcia_Cirac_Lewenstein_Zoller_variational}.
%
%
We could just as well use a Thomas-Fermi instead of Gaussian profile; however
the results are very similar and do not substantially affect any of the
arguments we make in this work.
%
%
The two parameters turn out to be not independent but simply related
($\beta(t) \propto\del_t\ln\sigma(t)$).  There is thus effectively a single
dynamical parameter describing the system, namely the cloud radius
$\sigma(t)$.  The equation of motion for $\sigma$ is
\begin{equation}
\sigma\frac{{\rm d}^2\sigma}{{\rm
    d}t^2}+\sigma^2-\frac{1}{\sigma^2}-\frac{U(t)}{(\sqrt{2\pi}\sigma)^D}=0 \, , 
\label{eq:variational_radiusEq} 
\end{equation}
and the energy is 
\begin{equation}
E[\sigma] =  
\frac{D}{4}\left[
\frac{1}{\sigma^2}+\sigma^2
+\left(\frac{{\rm d}\sigma}{{\rm d}t}\right)^2 \right]
~+~ \frac{U(t)}{2(\sqrt{2\pi}\sigma)^D}  \, .
\label{eq:variational_energyEq}
\end{equation}
The radius description based on the above two equations is suitable for
describing breathing-mode oscillations.  For constant $U$, trying
small-amplitude oscillatory solutions of form $\sigma =
R_0+\rho\sin(\Omega{t})$, Eq.\ \eqref{eq:variational_radiusEq} yields
$\Omega\sim\sqrt{D+2}$ for the breathing mode frequency at large $U$.  We also
find that $R_0$ satisfies the stationary equation for the radius, i.e., that
$R_0[U]{\sim}U^{1/(D+2)}$ in the Thomas-Fermi limit of large $U$.  Also,
Eq.\ \eqref{eq:variational_energyEq} shows that the excitation energy with a
breathing-mode oscillation of amplitude $\rho$ scales as ${\sim}\rho^2$ with
the oscillation amplitude $\rho$; we will use this for our energy analysis
below.

\begin{figure}
\centering
\includegraphics*[width=0.99\columnwidth]{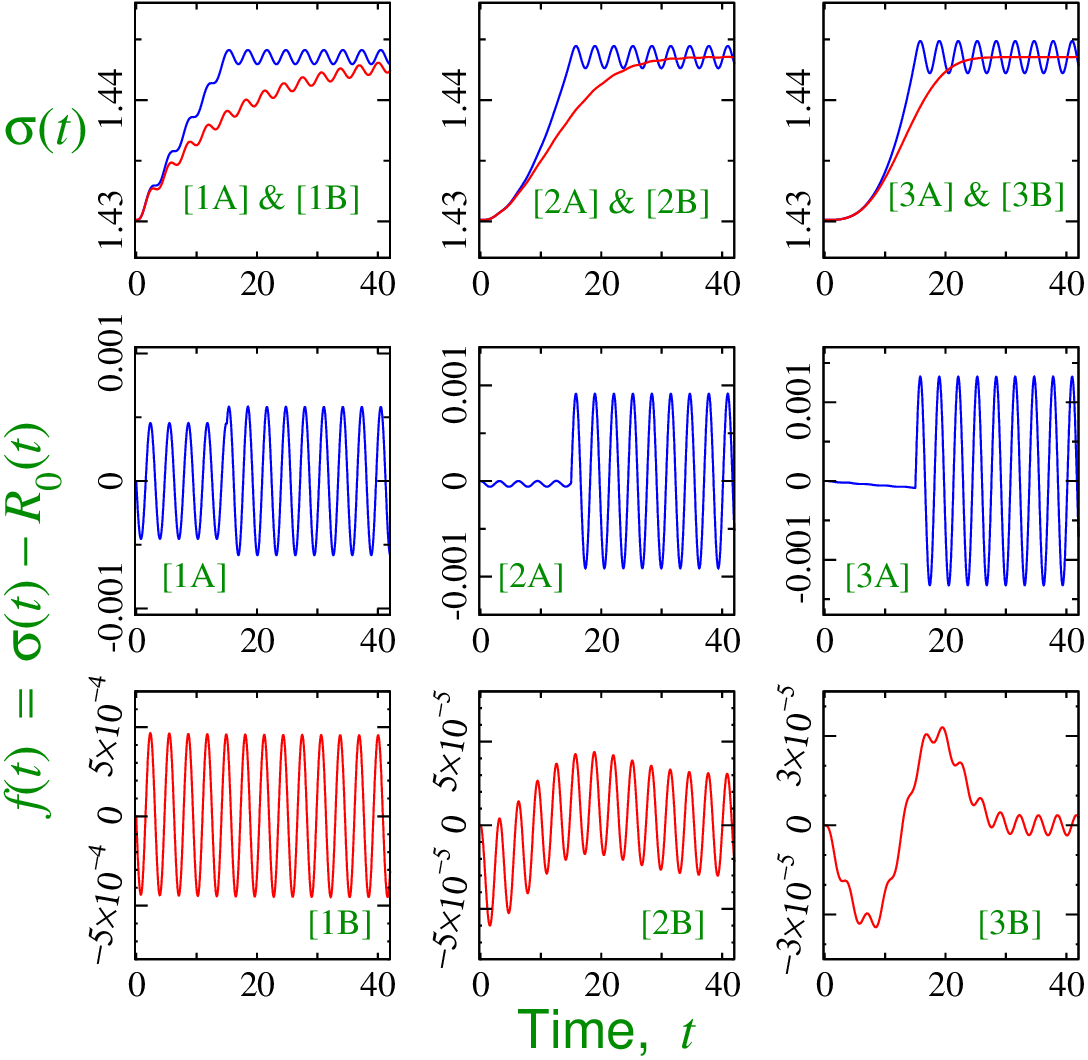}
\caption{ \label{fig_radusOscillatns}
Top row: condensate radius dynamics $\sigma(t)$ for various ramp shapes,
$\tau=15$.  Center and lower rows: deviation $f(t)$ from the `instantaneous'
ground-state radius $R_0(t)$.
Dimensions $D=2$, ($U_{i}$,$U_{f}$) = (20,21).
}
\end{figure}

\emph{Radius dynamics interpretation} ---
We now proceed to explain the $Q(\tau)$ behaviors presented above, in terms of
radius dynamics.  Figure \ref{fig_radusOscillatns} (top row) shows the radius
evolving as a function of time for various ramp shapes, for reasonably large
$\tau$.  In the center and bottom rows, we show the deviation of $\sigma(t)$
from the equilibrium radius corresponding to the instantaneous value of the
interaction, $R_0(t) = R_0[U(t)]$.  For a truly adiabatic ramp, $\sigma(t)$
would follow $R_0(t)$ exactly; therefore the deviation $f(t)=\sigma(t)-R_0(t)$
is at the heart of non-adiabaticity and the behavior of this quantity
determines the amount of heating.  After the ramp, the $f(t)$ function is
purely oscillatory; the heating scales as the square of the oscillation
magnitude.  Figure \ref{fig_radusOscillatns} presents radius dynamics for
$D=2$; the 1D and 3D cases are very similar.

For the [A] ramps with derivative discontinuities (middle row), the magnitude
of the final oscillations of $f(t)$ is determined at the ramp endpoint.  For
the $(t/\tau)^{\alpha}$ ramp, the oscillation magnitude is
$\ord(\tau^{-\alpha})$ during the ramp, and turns into $\ord(\tau^{-1})$ after
the endpoint kink.  For $\alpha>1$, the final $\ord(\tau^{-1})$ oscillation is
parametrically larger than the during-ramp $\ord(\tau^{-\alpha})$ oscillation.

We will first explain the $\sim\tau^{-1}$ scaling of oscillations initiated at
the kink.
If we neglect the smaller oscillations at $t<\tau$, the radius
$\sigma(t){\approx}R_0(t)$ at the kink $t=\tau$ has ``correct'' value for
$U=U_{f}$, i.e. $f$ is negligible.
%
However the derivative is nonzero, $\sigma'(t) {\approx} R_0'(t)|_{t=\tau}$,
which scales as $\sim\tau^{-1}$.  
Thus we have the following ``initial'' conditions at $t=\tau^+$ for subsequent
evolution: $f(\tau)=0$, $f'(\tau^+)=c_0\tau^{-1}$.
Using $f(t>\tau)\approx\rho\sin(\Omega{t}+\delta)$, these initial values imply
$\rho\sim\tau^{-1}$.  This explains the $\ord(\tau^{-1})$ oscillation
magnitude and hence $\ord(\tau^{-2})$ residual energy for ramps having a
derivative jump at the endpoint.

The oscillations of $Q(\tau)$ (Figure \ref{fig_excessEnergies}d) can be
explained by relaxing the approximation $f(t<\tau)\approx0$ made above. 
%
%
The
small oscillations of $f(t<\tau)$ guarantee that $\sigma'(t=\tau)$ oscillates
around $R_0'(t=\tau)$ as a function of $\tau$.  This results in the final
breathing mode amplitude $\rho$ to oscillate around its $\ord(\tau^{-1})$
value as a function of $\tau$.  Since the $f(t<\tau)$ breathing-mode strength
is smaller for larger $\alpha$, the oscillations of the heating with $\tau$
are weaker for larger $\alpha$, as seen in Figure \ref{fig_excessEnergies}d.

The lowest-row panels of Figure \ref{fig_radusOscillatns} focus on the smooth
[B] ramps.  In these cases, the breathing-mode strength ($\sim\tau^{-\alpha}$)
initiated at the beginning of the ramp remains unchanged; there is no kink to
abruptly create larger oscillations.
We therefore need only to explain the strength of oscillations at the
beginning of the ramp, where $r(t/\tau)=1-e^{-(t/\tau)^{\alpha}} \approx
(t/\tau)^{\alpha}$.
%

We first rewrite Eq.\ \eqref{eq:variational_radiusEq} as an equation for
$f(t)$.  To simplify notation, we will write this out explicitly only in the
Thomas-Fermi limit, $U_{i,f}\gg1$, and small oscillations, $f(t){\ll}R_0(t)$.
(The arguments can of course be modified to go beyond the Thomas-Fermi
restriction. Small $f(t)$ is guaranteed for large $\tau$.)  We obtain
\begin{equation}
f''(t) + \Omega^2f(t) + \frac{u''(t)}{(D+2)u^{\frac{D+1}{D+2}}} - \frac{(D+1)u'(t)^2}{(D+2)^2u^{\frac{2D+3}{D+2}}} = 0
,
\end{equation}
with $u=U/(2\pi)^{D/2}$.  The first two terms give pure oscillatory behavior
(breathing mode at fixed $u$); the last two terms are corrections due to
time-varying interaction.

We first treat ramps with zero initial slope, i.e., $\alpha>1$. The initial
conditions at $t=0^+$ are then $f(0)=f'(0)=0$.
With $u=u_{i}+ (\delta{u})(t/\tau)^{\alpha}$, the $u''$ correction is dominant
compared to the $u'^2$ correction at $t\ll\tau$.  The dominant correction
terms take the form $c_1/\tau^2$ for $\alpha=2$, and $c_1t/\tau^3$ for
$\alpha=3$.
%
%
The solutions of the resulting differential equation are linear combinations
of oscillatory trigonometric functions and algebraic functions.  It is
straighforward to verify that the boundary conditions $f(0)=f'(0)=0$ force the
oscillatory part to have coefficients scaling as $\sim\tau^{-2}$ for
$\alpha=2$ and $\sim\tau^{-3}$ for $\alpha=3$.  This explains the
$Q\sim\tau^{-2\alpha}$ behavior for integer $\alpha>1$.

The $\alpha=1$ case is slightly different.  The initial condition still
involves $\sigma'(0)=0$, but since $R_0(t)=[u(t)]^{1/(D+1)}$ has finite slope
at $t=0^+$, this now corresponds to $f'(0^+)=-R_0'(0^+)=-c_3/\tau$. This
initial condition leads to a purely oscillatory $f(t)$ with amplitude
$\sim\tau^{-1}$, which explains $Q(\tau)\sim\tau^{-2}$ for $\alpha=1$.

\begin{figure}
\centering
\includegraphics*[width=0.99\columnwidth]{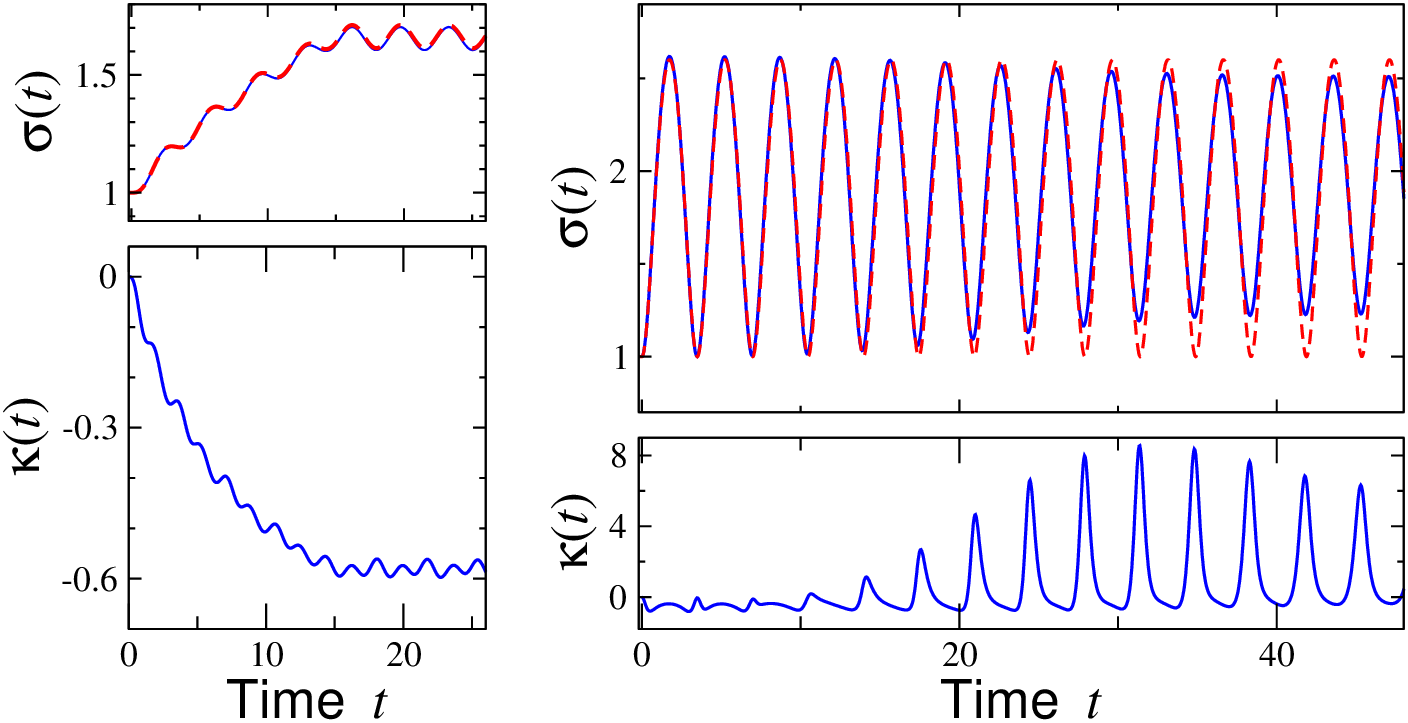}
\caption{ \label{fig_beyond}
Physics beyond pure radius description.
$D=1$, ($U_{i}$,$U_{f}$) = (0,10).
Left panels: $\tau=15$ ramp.  Right panels: instantaneous quench.  
On top panels, radius from full GP calculations (full lines) compared to
single-parameter variational calculations (dashed lines).
Lower panels show kurtosis, not accessible in the radius-only description.
}
\end{figure}

\emph{Beyond pure radius dynamics} ---
We justified our single-parameter analysis by noting that $Q(\tau)/Q(0)$ is
reproduced splendidly by such a description (Figure
\ref{fig_excessEnergies}c).  The GP dynamics is of course richer than this
minimal description, one indication of which is that the \emph{un-normalized}
heat function $Q(\tau)$ obtained from Eqs.\ \eqref{eq:variational_radiusEq},
\eqref{eq:variational_energyEq} deviates from full-GP results.
Although a complete study of all aspects of shape dynamics induced in a ramp
is beyond the scope of the present investigation, in Figure \ref{fig_beyond}
we show some basic additional effects.  After the size, the next obvious shape
characteric is the kurtosis $\kappa$, related to the fourth moment of a
distribution, such that a gaussian has $\kappa=0$ and a more ``rectangular''
(sharper-peaked) distribution has $\kappa<0$ ($\kappa>0$).
The left panels of Figure \ref{fig_beyond} show that in a slow ramp, the
kurtosis oscillates around the instantaneous equilibrium value just as the
radius does. 
In the right panels, the large-amplitude radius oscillations after a sudden
($\tau=0$) quench is seen to vary from the single-parameter description; there
are dramatic jumps of $\kappa(t)$ at the times when the $\sigma(t)$ deviation
is prominent.  The energy clearly leaks from breathing-mode oscillations into
other channels, even though the total energy is conserved.

\emph{Relation to other results; Open questions} ---
We have addressed the fundamental question of adiabaticity in the context of a
paradigm system of cold-atom physics, namely, 
a Bose condensate in a trap.
We find that the heat function $Q(\tau)$, which characterizes
non-adiabaticity, to have an overall power-law decay.
The general intuition is that the decay of $Q(\tau)$ should be
exponential if there are no gapless points in the path of the ramp; thus
finite systems are generically expected to have exponential decay,
e.g., Ref.~\cite{Venumadhav_PRB2010}.  Remarkably, the GP description
successfully mimics a gapless thermodynamic limit by providing power-law
decay of $Q(\tau)$, although concepts like gap or density of states are not
meaningful within the GP description.

Our result $Q(\tau)\sim\tau^{-2\alpha}$ for smooth ramps is consistent with
adiabatic perturbation theory:
if our ramps are put into the form $r(t){\sim}vt^\alpha/\alpha!$, we get
$Q(\tau){\sim}\tau^{-2\alpha}{\sim}v^2$, which is the generic perturbative
expectation \cite{Polovnikov_AdiabaticPertThy}.  In the formulation of
Ref.~\cite{EcksteinKollar_NJP2010} (Sec.~3), the $F(x)$ function for our ramp
can be shown to have asymptotic form $x^{-2\alpha}$, which translates to an
``extrinsic'' contribution $Q(\tau)\sim\tau^{-2\alpha}$.
%
The GP description thus retains nontrivial dynamical information pertaining to
the full quantum description, despite being ``merely'' a nonlinear
differential equation.


Another feature we have explored is the sensitivity to a final kink in the
ramp shape.  A recently discovered effect of such kinks is logarithmic
contributions to $Q(\tau)$ \cite{Polovnikov_AdiabaticPertThy,
  EcksteinKollar_NJP2010, DoraHaqueZarand_arxiv10}.  The effect we have found
(kink induces larger oscillations overwhelming initial excitation) is quite
different.  It is an open question whether or not this is unique to the
present system.

Oscillations of $Q(\tau)$ are relatively poorly understood, and may well be
generic in many-body ramps.  In our case, it appears explicity due to ramp
shape kinks.  Like other $Q(\tau)$ features, we have provided a very physical
interpretation in terms of radius oscillations.  In other known examples of
$Q(\tau)$ oscillations \cite{CanoviRossiniFazioSantoro_JSM09,
  EcksteinKollar_NJP2010, Venumadhav_PRB2010}, the physical explanation of
oscillations, where known, are all different.

The present work opens up several new research avenues.  Ramps in the trapping
frequency should also induce radius oscillations, but details may well be
different from interaction ramps.  Physical insights developed in our study of
$Q(\tau)$ can perhaps be applied to better understand ``optimal ramp'' studies
seeking to find ramp paths producing minimal heating \cite{Muga-JPB-2009}.
Another natural extension of our work is a treatment of condensates beyond the
GP description, e.g., truncated Wigner schemes or numerical full quantum
treatments of few-boson systems.


\bibliographystyle{amsplain}


\end{document}